\documentclass[article]{agujournal2019_arxiv}
\usepackage{url} 
\usepackage{lineno}
\usepackage{soul}

\usepackage{chngcntr}

\usepackage{graphicx}
\usepackage{amsmath}
\usepackage{mathtools}

\usepackage{natbib}
\pdfoutput=1

\newcommand{\vect}[1]{\boldsymbol{#1}}
\usepackage{mathtools,bbm}

\usepackage{amssymb}
\usepackage{gensymb}

\usepackage[nomarkers,tablesonly,figuresonly]{endfloat}

\usepackage{amsmath}

\usepackage{bbm}

\usepackage{float,graphicx}
\usepackage{tabularx, booktabs}

\usepackage{xcolor}

\usepackage{algorithm}
\usepackage[noend]{algpseudocode}

\usepackage{footnote}

\algrenewcommand\algorithmiccomment[1]{\hfill \(\phantom{ }\) #1}

\newcommand{\algorithmfootnote}[2][\footnotesize]{%
  \let\old@algocf@finish\@algocf@finish
  \def\@algocf@finish{\old@algocf@finish
    \leavevmode\rlap{\begin{minipage}{\linewidth}
    #1#2
    \end{minipage}}%
  }%
}

\draftfalse

\journalname{EPS}

\usepackage{hyperref}

\hypersetup{
    colorlinks,
    citecolor=black,
    linkcolor=black,
    breaklinks=true
}

\begin{document}

\title{Double Difference Earthquake Location \\ with Graph Neural Networks}

\authors{Ian W. McBrearty and Gregory C. Beroza}

\affiliation{1}{Department of Geophysics, 397 Panama Mall, Stanford University, Stanford, California 94305-2215, USA} \vspace{0.2 cm}

\correspondingauthor{I.W. McBrearty}{imcbrear@stanford.edu}

\begin{keypoints}
\item Graph Neural Networks
\item Double Difference Earthquake Location
\end{keypoints}

\begin{abstract}

Double difference earthquake relocation is an essential component of many earthquake catalog development workflows. This technique produces high-resolution relative relocations between events by minimizing differential measurements of the arrival times of waves from nearby sources, which highlights the resolution of faults and improves interpretation of seismic activity. The inverse problem is typically solved iteratively using conjugate-gradient minimization for up to $\sim$100,000's events, however the cost scales significantly with the total number of sources and stations considered. Here we propose a Graph Neural Network (GNN) based earthquake double-difference relocation framework, Graph Double Difference (GraphDD), that is trained to minimize the double-difference residuals of a catalog to locate earthquakes. Through batching and sampling the method can scale to arbitrarily large catalogs.  This is a distinct use of machine learning for earthquake catalog development because the network is trained to minimize the double-difference residuals over an entire catalog simultaneously, rather than for individual per-event predictions. Our architecture uses one graph to represent the stations, a second graph to represent the sources, and creates the Cartesian product graph between the two graphs to capture the relationships between the sources and stations (e.g., the residuals and travel time partial derivatives). This key feature allows a natural architecture that can be used to minimize the double-difference residuals. We implement our model on several distinct test cases including seismicity from northern California, T\"{u}rkiye, and northern Chile, which have highly variable data quality, and station and source distributions. We obtain high resolution relocations in these tests, and our model shows adaptability to variable types of loss functions and location objectives, including learning station corrections and mapping into the reference frame of a different catalog. Our results suggest that a GNN approach to double-difference relocation is a promising direction for scaling to very large catalogs and gaining new insights into the relocation problem. 

\end{abstract}

\section{Introduction}

There are well-established procedures to automate the mapping of seismograms observed on a network of seismic stations into an earthquake catalog, which is a list of source origin times, locations, magnitudes, and possibly source types (blast or earthquake) and source mechanisms (focal mechanism or moment tensors). The serial tasks of picking arrival times \citep{allen1978automatic}, association \citep{ringdal1989multi}, and location \citep{klein1978hypocenter} constitute a typical workflow to develop seismic event catalogs, and a variety of software algorithms exist to handle these steps (e.g., \citet{weber2007seiscomp3, friberg2010earthworm, zhang2022loc, zhu2023quakeflow}).

Despite many improvements to catalog construction workflows, catalogs inevitably contain both uncertainty and bias compared to the true seismic event history \citep{kagan2003accuracy, woessner2005assessing}.  For example, association and location routines are typically applied with 1D velocity models that are approximations to the true 3D Earth structure. There can be false detections, incorrectly associated picks, inaccurate pick times, and other errors corrupting the interpretation of the data \citep{lahr1994earthquake, di2006automatic}. Catalogs can be improved though post-processing techniques, such as re-calibrating the initial velocity model based on the detected events (e.g., \citet{kissling1994initial}), removing outliers \citep{zhu2021multi}, or estimating station corrections \citep{lomax2022high}, all of which can reduce location uncertainties and bias. 

One of the most important steps in developing a catalog is determining earthquake hypocenters. Initial treatments of this problem focused on locating individual events, but joint hypocentral location techniques \citep{douglas1967joint} that re-locate groups of seismic events with respect to each other by treating the entire catalog as a combined entity (rather than treating each earthquake separately) have a distinct advantage because they can reduce location uncertainty and bias. The HypoDD double difference earthquake relocation algorithm \citep{waldhauser2000double} is a joint hypocentral determination method that directly exploits differential arrival time measurements to suppress the common mode error that results from unmodeled path effects. In doing so HypoDD focuses sensitivity on the near source region, and can constrain the positions of sources with greater precision and accuracy. 
 
 \subsection{Double Difference Location}
There is strong empirical evidence supporting the effectiveness of HypoDD in the clarity it provides in resolving complex fault structures as illuminated by seismicity. The theoretical underpinnings of how, why, and where double-difference methods work have also been explored by \citet{wolfe2002mathematics, menke2022differential}, who show the amount of improvement enabled by differential measurements is limited by the uncertainty of the velocity model and the correlation structure of the residuals. Improved location precision is also possible if relative arrival time measurements are made using waveform-based cross-correlation (e.g., \citet{poupinet1984monitoring, rubin1999streaks, schaff2004optimizing, waldhauser2008large, hauksson2012waveform, shelly2020high}). Double-difference location with cross-correlation based arrival time measurements is widely considered as the most effective way to produce high resolution locations, and hence, optimizing its use and application is an important goal for earthquake catalog development. 

Despite HypoDD's great success, challenges remain in utilizing the technique, and several extensions to the initial method have been developed over time to address some of them \citep{zhang2003double, trugman2017growclust, lin2018source}. Due to the direct least-squares formulation of the base objective function in HypoDD (minimizing the `double-difference' residuals), the inverse problem grows substantially and poses memory challenges for large catalogs \citep{waldhauser2008large}. For large data sets, HypoDD uses the linearized conjugate gradient descent algorithm, but this can introduce convergence and instability challenges \citep{hager2006survey}, does not yield a model covariance matrix for error characterization, and is still subject to eventual limitations on the number of earthquakes that can be relocated simultaneously (e.g., $\sim$100,000's events). One option to address these challenges is to build the catalog iteratively, and to re-locate newly detected events against a fixed set of reference events \citep{waldhauser2000double}. GrowClust  \citep{trugman2017growclust} takes an alternative approach of splitting the initial catalog into core spatial clusters, followed by a double-difference least squares optimization within each cluster. Another approach is to apply HypoDD over many disjoint or slightly overlapping spatial windows in parallel (e.g. \citet{waldhauser2008large, yano2017japan}). This approach comes with the additional complication of resolving locations in overlapping regions and near window boundaries. Yet another approach is to use both the standard double-differences between nearby sources, for a fixed station, and double-differences between nearby stations, for a fixed source (e.g., \citet{yuan2016double}). Some extensions of the method also allow 3D velocity models \citep{trugman2023growclust3d}, and TomoDD \citep{zhang2003double} allows simultaneous optimization of the 3D velocity model and source locations.

All of these techniques require substantial hyper-parameter tuning, which introduces complex tradeoffs in performance based on the values chosen \citep{yu2024location}. Of particular note is the ambiguity of which pairs of sources to compare in the double-difference routine, as that impacts the size of the inverse problem, the computational resources required, and controls the landscape of the objective function. HypoDD, GrowClust, and TomoDD each use heuristic algorithms to decide which pairs of sources (for a fixed station) to compare, as well as the weighting to apply to different data points, while accounting for factors that influence the results such as: the maximum source-pair distances, minimum number of nearby sources, nearest neighbor distributions, and azimuthal coverage. Because the pairs of earthquakes being compared are updated iteratively, locations will change and some event pairs may separate too far from one another and be removed, while other event pairs may move closer to one another and be added \citep{zhang2003double}. The weighting between absolute travel time data and double difference data, and between initial pick times and cross-correlation based differential time measurements, can also be varied during successive iterations, for example by gradually increasing the weight given to cross-correlation differential time measurements for later iterations as locations converge \citep{waldhauser2000double}. Additionally, some algorithms remove pairs that have high residuals as probable outliers. Based on this variety of choices and challenges, and due to the recent success of machine learning (ML) in improving diverse seismological processing tasks \citep{mousavi2022deep}, we explore the potential of deep learning (DL) applied to the double difference relocation problem.

\subsection{The Case for Deep Learning}
An important advantage of a DL approach for double-difference location is that it is possible to apply the method to arbitrarily large seismic catalogs without memory issues. It is also possible that the DL solution will be less sensitive to user specified hyper-parameter choices, and more robust to outliers and noisy data. We note that double-difference relocation is fundamentally different than previous DL applications in seismology such as phase picking \citep{ross2018generalized, zhu2019phasenet}, location \citep{mousavi2019bayesian, van2020automated}, association \citep{ross2019phaselink, mcbrearty2023earthquake}, magnitude determination \citep{mousavi2020machine, dybing2024rapid}, polarity measurement \citep{ross2018p}, and back-azimuth estimation \citep{dickey2021baznet, kohler2023arraynet}, in that the solution is a joint prediction of the relative source positions of a potentially very large number of earthquakes simultaneously, rather than a single prediction of one (or a few) earthquakes at a time. Further, because earthquakes in the population will be observed by a different set of stations, and station travel time differences are the core input to the double-difference objective function, the data structure being processed contains relative information between many different pairs of stations and sources simultaneously. For this reason it is not straightforward to formalize the training problem or to choose the optimal architecture using most existing neural network models. 

We propose a deep learning structure that naturally identifies the key interaction in the data between `nearby source coordinates', for a fixed station, and `nearby stations', for a fixed source, while allowing for the heterogeneous set of a different number (and variably quality) of picks for each event. Our approach is based on a graph neural network that uses a Cartesian product graph, where one graph represents the stations, the other graph the sources, and the product captures the interactions between the two graphs (see graphical abstract). This is analogous to an approach that used a Cartesian product graph between stations and sources in a graph neural network to solve the phase association problem \citep{mcbrearty2023earthquake}. The graph representation allows the neural network to handle the heterogeneous distribution of sources and observing stations naturally, and simultaneously, while accounting for varying geometric configurations of stations and sources. Further, our model uses a travel time neural network surrogate (applicable to 1D or 3D models) that when combined with automatic differentiation through both neural networks allows direct optimization of the double-difference residual iteratively, while training the GNN and correctly modeling absolute travel times between sources and receivers. This transforms the problem of minimizing the double-difference residuals of a catalog into the iterative training of a deep neural network, which allows trading off memory cost for computation time.  An added benefit is that it can improve solution stability and reduce sensitivity to user-specified hyperparameters. 

\section{Methods}

\subsection{Overview}

We construct a graph neural network that accepts a station set, a sample from an earthquake catalog, initial source locations, input velocity model, and predicts the relative source offsets that will minimize the double-difference residual. Our approach must account for several challenges: (i) we cannot input the entire catalog at once, but must pass in instances (samples) from the catalog; (ii) the input data is multi-modal, as both source- and station- specific information must be processed simultaneously; and (iii) the choice of which pairs of sources, or pairs of stations to include in the double difference residual, must be specified. In the following, we describe our strategies for addressing these challenges, and in particular, the core design choice of our GNN, which utilizes the Cartesian product graph between stations and sources \citep{mcbrearty2023earthquake}, and which can naturally handle the pairwise nature of the input data.  We then demonstrate applications of the trained model to both real and synthetic data.

\subsection{Double-Difference Objective}

The key idea in the double-difference method \citep{waldhauser2000double} is that the relative arrival time of pairs of nearby sources measured on the same station are sensitive to source location differences, but are relatively insensitive to long-range path effect uncertainties. This occurs because they share nearly the same ray path, and hence the `common mode' uncertainty largely cancels from the differential arrival time measurements \citep{menke2022differential}. When a joint population of relative earthquake relocation vectors is estimated to model the arrival-time differences between many pairs of stations and sources simultaneously, the optimal relocation vectors refine the locations of earthquakes, and these tend to coalesce onto active faults or into known quarries \citep{waldhauser2008large}.

Following \citep{waldhauser2000double}, the double difference residual between two events $i$ and $j$, recorded on station $\vect{s}$, is given by

\begin{equation} \label{eq: double difference}
\begin{split}
    d_{ij} &= (\tau_{i} - \tau_{j})^{obs} - (\tau_{i} - \tau_{j})^{cal} 
\end{split}
\end{equation}

\noindent where $\tau_{i}^{obs}$ and $\tau_{i}^{cal}$ are the observed, and modeled arrival times for the $i$th event. Since all earthquakes in the catalog have an initial location ($\vect{x}_i$), and we are seeking a perturbation (or update) to this initial location ($\tilde{\vect{x}}_i = \vect{x}_i + \delta\vect{x}_i$), \eqref{eq: double difference} can also 
be re-written as:

\begin{equation} \label{eq: double difference residual 2}
\begin{split}
    d_{ij} &= (\tau_{i} - \tau_{j})^{obs} - \Bigl(T_{\vect{s}}(\tilde{\vect{x}}_i) - T_{\vect{s}}(\tilde{\vect{x}}_j) \Bigr) \\
    &= \Bigl(\tau_{i}^{obs} - T_{\vect{s}}(\vect{x}_i)\Bigr) - \Bigl(\tau_{j}^{obs} - T_{\vect{s}}(\vect{x}_j)\Bigr) - \Bigl(T_{\vect{s}}(\tilde{\vect{x}}_i) - T_{\vect{s}}(\vect{x}_i)\Bigr) + \Bigl(T_{\vect{s}}(\tilde{\vect{x}}_j) - T_{\vect{s}}(\vect{x}_j)\Bigr) \\
\end{split}
\end{equation}

\noindent where $T_{\vect{s}}(\vect{x})$ represents the travel time to station $\vect{s}$ for source $\vect{x}$. The advantage of \eqref{eq: double difference residual 2} is that it isolates the initial observed residuals, $r_i = \tau_i^{obs} - T_{\vect{s}}(\vect{x}_i)$ to the left hand side, and the perturbations of theoretical arrivals, $\delta T_{\vect{s}}(\tilde{\vect{x}}_i, \vect{x}_i) = T_{\vect{s}}(\tilde{\vect{x}}_i) - T_{\vect{s}}(\vect{x}_i)$, to the right hand side. The double difference residuals $d_{ij}$ can then be written as

\begin{equation} \label{eq: loss function}
    d_{ij} = (r_i - r_j) - \Bigl(\delta T_{\vect{s}}(\tilde{\vect{x}}_i, \vect{x}_i) - \delta T_{\vect{s}}(\tilde{\vect{x}}_j, \vect{x}_j)\Bigr)
\end{equation}

\noindent where the goal is to minimize $d_{ij}$ by obtaining the optimal source relocations, $\tilde{\vect{x}_i}$, or equivalently, perturbations, $\delta \vect{x}_i$, for each source. 

We train our model to predict $\delta \vect{x}_i$ for all sources in each input graph, and train it to minimize the $L1$ misfit between $r_i - r_j$ and $\delta T_{\vect{s}}(\tilde{\vect{x}}_i, \vect{x}_i) - \delta T_{\vect{s}}(\tilde{\vect{x}}_j, \vect{x}_j)$. We use the $L1$ norm as the loss function to enhance robustness to outliers. Our approach requires that $T$ is differentiable with respect to $\vect{x}$, for fixed $\vect{s}$, since we will differentiate the loss function through the travel time calculator, $T_{\vect{s}}(\tilde{\vect{x}})$, for predicted relocation positions, $\tilde{\vect{x}}$. This is straightforward using pre-trained neural network surrogate travel time calculators for a chosen velocity model \citep{smith2020eikonet, anderson2023emulation}, which are automatically differentiable. Finally, we note that origin time perturbations of sources $\delta t_i$ simply correspond to a constant time shift in theoretical arrival times for all stations, and that the loss is computed in parallel for both P-and S-waves (or other phase types) separately.

\subsection{Absolute and Calibration Objective}

Minimizing the double difference objective can lead to a drift in source locations of isolated clusters. Typically, this issue is suppressed by applying damping to the mean source re-location offsets \citep{waldhauser2000double}. To obtain a similar regularization effect, we include the option for an additional loss related to the absolute travel time residuals (Eq. 4), though we down-weight this loss significantly compared to the double difference loss (e.g., 1:4). Incorporating this loss is straightforward, as it requires the same inputs, training data, model, and outputs ($\delta \vect{x}_i$) we have already considered. This additional loss is not required, but can stabilize and reduce the non-uniqueness in the predicted locations, as shown in our results.

In some cases there is an added challenge that the initial event catalog is built with a sub-optimal velocity model. Lateral velocity contrasts across shallow faults, for example, can lead to event locations that are displaced from their true locations by several kilometers \citep{kim2016seismicity}. Station corrections can partially counteract the effect of unmodeled velocity model structure \citep{lomax2022high}, and reduce spatial bias in event locations; however, joint inversion of source locations and station/path corrections is a nuanced problem \citep{pujol1988comments, richards2000earthquake} due to variations in convergence rates, station coverage, data quality, initial velocity models, and complexity in crustal structure.

We choose to account for these challenges by introducing a few additional loss terms to the model. Specifically, we estimate the average travel time residual for each station of each input graph, and train the model to predict station corrections that minimize this residual (Eq. 4c). To facilitate easier comparisons between the initial and reference catalogs, we add a calibration loss that promotes re-locating our events more closely to events we intend to match in a reference catalog. This can be helpful when the two catalogs being compared were built with different velocity models, or if one catalog uses station corrections and the other does not. This issue can arise frequently in practice, and complicates direct catalog comparisons. The calibration loss operates on a chosen subset of events of the input catalog that are matched to a reference catalog (e.g., events occurring in a similar spatial and temporal window).  It adds an additional loss to penalize the misfit between the observed arrival times and the theoretical arrival times predicted from the reference hypocenters (Eqs. 4a,b) while also incorporating station corrections. 

That is, in addition to the double difference loss, $\mathcal{L}_{diff}$, for each station and all picks it has with `matched' earthquakes with hypocenter $\vect{x}_r$, we compute the additional losses of


\begin{subequations}
\begin{align}
    \mathcal{L}_{cal} &= \sum_{\vect{s} \in \mathcal{S}} \sum_{i \in M_{\vect{s}}} \lvert \tau_i^{obs} - (T_{\vect{s}}(\vect{x}_r) + \Delta \vect{s} ) \rvert \\
    \mathcal{L}_{abs} &= \sum_{\vect{s} \in \mathcal{S}} \sum_{i \in E_{\vect{s}}} \lvert \tau_i^{obs} - (T_{\vect{s}}(\tilde{\vect{x}}_i) + \Delta \vect{s} ) \rvert \\
    \mathcal{L}_{sta} &= \sum_{i \in E_{\vect{s}}} \lvert \tau_i^{obs} - (T_{\vect{s}}(\tilde{\vect{x}}_i) + \Delta \vect{s} ) \rvert
\end{align}
\end{subequations}

\noindent where $\Delta \vect{s}$ represents the predicted station corrections for station $\vect{s}$ of the station set $\mathcal{S}$, and sets $E_{\vect{s}}$ and $M_{\vect{s}}$ represent the event indices that station $\vect{s}$ has an associated pick for, and the subset of these indices that also have a matched earthquake in the reference catalog, respectively. Hence, (4a) promotes training to predict station corrections that accurately model the observed arrival times while using the reference catalog source coordinates for the subset of matched events. The term (4b) promotes learning re-location predictions of source locations, $\tilde{\vect{x}}_i$, that when combined with the predicted station corrections also model the observed arrival times for all events in the catalog. By inspection, the calibration loss (4a,b) should guide the relocation predictions, $\tilde{\vect{x}}$, to become consistent with the reference positions, $\vect{x}_r$, as this is the only way both losses can be minimized simultaneously. The (4c) loss is different than the others in that $\mathcal{L}_{sta}$ is computed separately for each station, so that unlike computing a bulk residual over the entire catalog as in (4a,b), this loss promotes locations and station corrections that yield a mean average residual near zero for each station, which is a primary objective of incorporating station corrections. The combination of these losses in addition to the double difference loss $\mathcal{L}_{diff}$ effectively reduces spatial bias due to poorly modeled velocity structure and enables the relocated catalog to more closely conform with a reference catalog.

\subsection{Cartesian Product Graph}

The input to our GNN is a graph defined by a product of two other graphs: the source graph, $\mathcal{X}$, that links nearby sources, and the station graph, $\mathcal{S}$, that links nearby stations (see graphical abstract). This product graph has the unique property that each node of the product represents a unique pair of source and station nodes (and hence each node represents an associated pick), and the edges connecting different nodes in the product are determined by the edge structure defined by either component graph, $\mathcal{X}$, and $\mathcal{S}$, separately. These graphs are typically very large (e.g., $>$100,000's of nodes; Table 1), but can be made very sparse, due to having highly distributed interactions resulting in effective scaling and stability behavior. This type of graph has been effectively used in past GNNs for seismic phase association \citep{mcbrearty2023earthquake}, and for location and magnitude estimation \citep{mcbrearty2022earthquake}. The strength of this approach is that the station data are analyzed from the perspective of each source node (model-space node), simultaneously. Graph convolutions transfer information along both connected stations and connected sources to extract higher order features and insights related to the coherence and similarity (or dissimilarity) of observations over variable physical scales, stations, and regions. It also incorporates geometric data such as Cartesian offset vectors and distances between connected nodes during graph convolution, so it can learn to treat the observed features differently depending on the relative station and source distances and distributions. Multiple rounds of graph convolution allow the extraction of latent features that combine and distill information over larger physical apertures \citep{battaglia2018relational}, which can promote robustness to noisy data and internal consistency of the feature space.

Our use of the Cartesian product differs somewhat from that used in \citep{mcbrearty2023earthquake}. Specifically, in \citet{mcbrearty2023earthquake}, all pairs of sources and stations were considered, and the source nodes acted as monitoring nodes to detect travel time moveout curves from continuous pick data, which required only $\sim$100s$-$1000s of uniformly spatially distributed nodes. Here, the source nodes represent individual events, so their distribution is heterogeneous, all events must be considered, and there may be millions of them. Hence, it is computationally infeasible to construct the full Cartesian product graph between sources and stations, as done in \citet{mcbrearty2023earthquake}. Instead, we construct a series of (randomly generated) subgraphs of the full Cartesian product, where we only consider large subsets of source and station pairs from the full Cartesian product at a time. For each source, we only consider the subset of stations that have an associated pick for that source, which can be far fewer than the total number of stations. Despite being randomly generated, these subgraphs retain much of the same properties as the full Cartesian product, and are still highly sparse, locally well-connected, and large, such that they offer an effective structure to process data through graph convolutions. Similar subgraph sampling techniques are widely used for scaling GNN applications to very large graphs \citep{alsentzer2020subgraph, frasca2022understanding}. In the following we describe our technique for selecting a randomly generated subgraph from the Cartesian product graph that maintains a spatially coherent set of sources for each input.

\subsection{Turning Catalogs into Input Graphs}

To limit the total memory cost we construct the input catalog graphs to the GNN by sampling a subset of sources from the full catalog. While doing so, we must sample a sufficient number of nearby sources so that nearby sources can interact while being processed to constrain their relative locations. By sampling many different random instances of input graphs and training the GNN over a large distribution of possible inputs, we obtain a GNN that can be applied to any of the inputs despite their diversity in source locations, monitoring stations, relative source distributions, and number of sources. After training, we can apply the model over many instances of the input graphs and obtain the average relocation prediction for all sources. While constructing these inputs we must decide which picks (or stations) to include for each source, and which edges to add between pairs of nearby sources and nearby stations. Our scheme for constructing these graphs and sampling sources is a simple heuristic scheme that requires minimal adjustment of hyperparameters depending on the physical scale of the region and the total size of the catalog used. 

The basic idea of the approach is to select a small number of sources as seed nodes, then add many random sources within a nearby radius of these sources, and an additional random sampling of sources from a larger neighborhood around these additional sources. Then, all associated stations for each selected source are included, and many edges are added between both nearby stations, and nearby sources. This typically results in input samples with $\sim$10,000 random sources, $\sim$100,000 total source-station pairs, and millions of edges (Table 1), which ensures each input graph is highly random and distributed over the full catalog, although with each source still having many nearby neighbors. Additional details on this graph construction scheme are given in the Supplementary Materials (Text S1).

For each input graph, we must decide on the input features that are defined for each input node. We choose to use an input feature that captures information related to both the input picks, their residuals, as well as information related to each source-station pair and individual source. Specifically, we define the input feature for each source-station pair ($\vect{x}$, $\vect{s}$) as

\begin{equation}
    \text{Input}(\vect{x}, \vect{s}) = [r_p,\hspace{0.1 cm} r_s,\hspace{0.1 cm} \partial_{\vect{x}} T_{\vect{s}}^p(\vect{x}),\hspace{0.1 cm} \partial_{\vect{x}} T_{\vect{s}}^s(\vect{x}),\hspace{0.1 cm} \vect{x} - \vect{s},\hspace{0.1 cm} || \vect{x} - \vect{s} ||,\hspace{0.1 cm} Deg(\vect{x}),\hspace{0.1 cm} M_p,\hspace{0.1 cm} M_s] 
\end{equation}

\noindent where we include the residuals, $r_p$, $r_s$, the travel time partial derivatives, $\partial_{\vect{x}} T_{\vect{s}}^p(\vect{x})$, $\partial_{\vect{x}} T_{\vect{s}}^s(\vect{x})$, the source-receiver offset vector and distance, $\vect{x} - \vect{s}$, $|| \vect{x} - \vect{s} ||$, the number of associated stations for each source, $Deg(\vect{x})$, and a binary scalar $M_p$ and $M_s$ indicating if a given source-station pair has an associated P or S pick, respectively. For clarity, we only include a source-station pair if there is at least one associated P or S wave (or both), so all entries in $\text{Input}(\vect{x}, \vect{s})$ will have at least one non-zero value of $M_p$ or $M_s$. When calibration losses are included in training (Eq. 4), the absolute source positions are also included in $\text{Input}(\vect{x}, \vect{s})$. This input (Eq. 5) effectively captures a summary of useful information regarding each source-station (or pick) pair, and hence is an effective input for enabling the GNN to minimize the double-difference residuals in training.

\subsection{Graph Neural Network Architecture}

To map the input features observed on the Cartesian product graph (Eq. 5), which include the travel time residuals, $r_p$, and $r_s$, for each pick, and the partial derivatives of the travel times, we first perform ten rounds of graph convolution directly on $\mathcal{S} \times \mathcal{X}$, followed by a bipartite graph convolution layer, which maps the Cartesian product graph into the source graph, $\mathcal{S} \times \mathcal{X} \rightarrow \mathcal{X}$, and yields the predicted offset vectors, $\delta \vect{x}$, for each source. This structure is analogous to the first two graph convolution modules used in \citep{mcbrearty2023earthquake}, referred to as the `Data Aggregation', and `Bipartite Read-In' layers, respectively. In the first module, the Data Aggregation graph convolution layers on $\mathcal{S} \times \mathcal{X}$ allow `mixing' of relevant information between both subsets of nearby sources and nearby stations, simultaneously, while accounting for the observed residuals of picks, travel time partial derivatives, and spatial features such as the relative offsets and positions of sources and stations. This allows the GNN to learn how the residuals on individual sources relate to the data observed on other nearby sources, while weighting the importance of neighboring data and transforming the input features into a rich latent vector recorded on each source-station pair. Since many rounds of graph convolution are repeated at this step, the GNN can extract a nuanced view of the catalog data that involves both nearby, and more distant sources and stations, simultaneously. 

After this initial data transformation, the Bipartite Read-In layer maps the latent features into a single summary latent vector for each source, by aggregating the feature vectors over all associated stations for that source by taking the mean, while also accounting for the source-station distances in this layer. This operation merges all the pick data for each source into a single aggregate feature vector for that source that summarizes information both specific to that source, and from other sources and picks in the catalog. This per-source latent feature vector is mapped through a fully connected neural network (FCN) to yield the final prediction of the source offset vectors, $\delta \vect{x}$, for each source in the input graph. When station corrections are also predicted, we embed the features from the Cartesian product graph into the station graph in addition to the source graph to obtain per-station correction predictions, $\Delta \vect{s}$. Through this construction, the GNN uses a natural architecture and information pathway that is suitable for predicting offset vectors $\delta \vect{x}$ that minimize the double difference residual. The relative residual information that is inherent to the double difference residual is directly encoded in the input, and the graph convolutions allow sharing of information between connected source pairs such that the architecture is well suited for this problem.

\subsection{Model Training}

After constructing the input graphs (Eq. 5) from the full catalog, we train the model to minimize the double difference residual. We achieve this by mapping the inputs through the GNN and obtaining per-source relocation offset vector predictions, $\delta \vect{x}$, passing these through the differentiable travel time calculator, $T$, and then computing the loss. We can use the double difference residual alone (Eq. 3), or include absolute travel times and calibration losses (Eq. 4) with a weighting of 4:1, respectively, depending on the use case. We use automatic differentiation of the L1 norm of the loss to update the GNN weights and minimize the residual, and weight S-waves by 1/2 compared to P-waves. We train on datasets of 50,000 randomly generated input graphs with a batch size of $\sim$5. All catalogs are initially filtered to only include events with at least ten neighbors within 10 km, which typically results in removing 5$-$15$\%$ of the initial sources. We implement the model in PyTorch Geometric \citep{fey2019fast} and use the Adam optimizer \citep{kingma2014adam} with a learning rate of 10$^{-3}$. In all tested cases, we see a gradual and smooth reduction in the double difference residuals that converge asymptotically (Fig. 1).

\section{Results}

\subsection{Applications}

To highlight the potential of our model we evaluate it under several real-world scenarios and on a synthetic test. These tests include applying the method on a deep learning enhanced catalog in northern California with $>$500 stations, the very active and spatially extended aftershock sequence of the 2023 T\"{u}rkiye Kahramanmara{\c{s}} earthquake that included the $M_w$ 7.8 and $M_w$ 7.6 mainshocks \citep{becker2024performance}, and a long-term, noisy catalog of the northern Chile subduction zone of $\sim$750,000 events that spans the great $M_w$ 8.2 Iquique earthquake \citep{mcbrearty2019earthquake}. These catalogs present a challenging test because they have $\sim$4$-$8$\times$ the number of events as routine catalogs, and include a large number of low magnitude events with potentially noisy picks, uncertain associations, and are observed under diverse real world monitoring conditions. These conditions represent a challenging test of the ability of the GNN-based double-difference relocation, Graph Double Difference (GraphDD). All the double-difference measurements we use are based on the original pick times, rather than cross-correlation, so location precision could be further improved with that additional processing. Finally, we use a synthetic test to verify that the method can process a very large ($>$1,000,000 event) catalog across northern California.

\subsection{Northern California: Application to a Regional Network Spanning a Transform Boundary}

We first apply our method to data from a combination of PhaseNet picking \citep{zhu2019phasenet} and GENIE association \citep{mcbrearty2023earthquake} in the same region as the initial application of GENIE. This catalog has $\sim$4$\times$ the number of events as the routine NCEDC catalog over the same time period and makes use of $>$500 stations throughout northern California with variable station density (including the NC, BK, BG, and BP seismic networks; Fig. S1). The station distribution is heterogeneous with stations concentrated around the Bay Area and dense sub-networks at Geysers (BG network) and Parkfield (BP network). The initial catalog events from the association output largely cluster around the major active faults (Figs. 2a,c); however, there is considerable scatter in the locations. 

After assembling the input graphs as described in the Methods section (Table 1) and training the GNN model for $\sim$50,000 steps with the double-difference (Eq. 3), absolute, and calibration objectives (Eq. 4), we see a notable and steady decrease in the loss function (Fig. 1), and a coalescence of the locations onto known active faults (Figs. 2b,d) with improved consistency with the reference NCEDC catalog (Fig. 3). Most obviously, we see a strong localization of seismicity onto the creeping sections of the San Andreas and Calaveras Faults, as well as the Sargent Fault, and a pronounced coalescence of events into many sub-clusters at Geysers (Fig. 4). In depth, we establish a vertical alignment of events along the Calaveras Fault (Fig. 5c,d). The depths at Geysers have better agreement with the NCEDC catalog after relocation (Fig. 5a,b), despite the large initial offset (e.g., $\sim$5 km) caused by differences in the applied velocity models. For earthquakes that are matched between our catalog and the NCEDC, the initial latitude and longitude standard deviations between matched events decrease from $\sim$0.019$^{\circ}$ to $\sim$0.013$^{\circ}$, the mean depth residual decreases from $\sim$4 km to 0.5 km, and the mean origin time residual decreases from -0.53 s to 0.04 s (Table 3). This highlights the reduction in bias compared to the reference catalog, along with the improvement of resolution and precision of locations.

This exercise affords some insight into the trade-offs of different training parameter approaches. Specifically, we find that the relocation results for the 30-nearest-neighbor criteria vs. the 80-km $\epsilon$-radius criteria to construct the station graphs are highly similar (Fig. S4). The 80 km-radius graphs typically include more edges for each station and hence typically have a broader epicentral distance range between stations, while the 30-nearest-neighbor graphs often have more local connections and are more sparse (Fig. S1). Despite this difference, only a slightly improved resolution of small clusters was obtained using the 80 km-radius graphs, and the locations are highly similar overall. This indicates that the obtained locations are robust with respect to the details of the graph construction schemes used, which is typical for GNN methods \citep{zhou2020graph}.

We tested training our model in two separate relocation updates, or `iteration steps'. That is, after training the model once and obtaining the estimated relocation updates for all events, we re-assemble the training data graphs for the new event positions, and re-train the model a second time and relocate the events again. We find that after the second round of re-training (or the `second iteration'), the relocation predictions remain highly consistent with the results of the first iteration (Fig. S5), although there is a slight improvement in resolution on some faults. The `density factor' of the catalog, which measures the average concentration of events (e.g., Table 2), increases by $\sim$10$\%$ between the first and second iterations, and the average double difference residual decreases a further 0.01 s compared to the initial relocation decrease from 0.18 s to 0.13 s. The predicted relocation offsets during the second iteration are also on average only $\sim$10$-$15$\%$ of the initial relocation distances (Figs. S6, S7). For either catalog, the standard deviations of individual events obtained from averaging relocation predictions over many different input graphs are typically low (e.g., $\sim$0.1$-$0.5 km), but can be large ($>$2 km) for seismicity where it is sparse. There are also systematically high values near Geysers and Parkfield, which may reflect uncertainty in the initial catalogs due to the increased proportion of small events in these areas. The systematic trends in standard deviation persist between both iterations of training, however the values decrease substantially during the second iteration (Fig. S5).

\subsection{Kahramanmara{\c{s}}: Application to a Geometrically Complex Aftershock Sequence}

On February 6th, 2023, the large Kahramanmara{\c{s}} $M_w$ 7.8 and Elbistan $M_w$ 7.6 earthquakes ruptured in T\"{u}rkiye and northern Syria, causing widespread destruction and an intense aftershock sequence. We apply our relocation method to the first five days of this aftershock sequence using an initial DL-enhanced catalog for this region \citep{becker2024performance}. This catalog was built using a combination of PhaseNet picking \citep{zhu2019phasenet} followed by GENIE association \citep{mcbrearty2023earthquake}, similar to the approach used for northern California, and contains $\sim$5$\times$ more events than the reference AFAD catalog \citep{AFAD2023catalog}. The seismicity included $\sim$3000 events/day during this interval and was distributed across multiple geometrically complex fault segments, including the East Anatolian Fault Zone (EAFZ), the \c{C}ardak fault, and the Narl{\i} splay fault that ruptured during the two main shocks (Fig. 6).

Here we train only to minimize the double-difference residual (Eq. 3). As in the northern California application, the model trains with a smooth loss that gradually decreases the loss function (Fig. 1) and simultaneously concentrates seismicity along the major fault traces (Fig. 6b). After relocation, the seismicity more clearly defines the EAFZ and the sub-parallel Narl{\i} splay fault that hosted the $M_w$ 7.8 initiation \citep{barbot2023slip}, and improves resolution along the \c{C}ardak fault. The depths we determine are more consistent with those in the AFAD catalog (Fig. 6c; Table 3), and the overall spatial pattern of events is similar (Fig. 6d), despite the fact we did not include the calibration or absolute loss function objectives (Eq. 4). During training, the `density factor' increased by $\sim$50$\%$, and the average double-difference residual decreased from 0.90 s to 0.68 s. Interestingly, the average standard deviation of relocation vector distances was comparable to the average offset distance length (e.g., 5.7 km and 6.2 km, respectively; Table 2), unlike our results for California where the standard deviations were $\sim$1/10th the average offset length.

To test the sensitivity of the relocation result to the combination of terms used in the loss function, we trained the T\"{u}rkiye model separately for only the double-difference loss (Fig. 7a), the double-difference loss and absolute loss (Fig. 7b), and the double-difference loss with absolute loss and station corrections (Fig. 7c). We also tested training for the double-difference loss using the `memory state' mechanism variant of the model (Fig. 7d; Supplemental Text S2), which passes the previous predictions of relocation values back into the model as additional inputs. All four relocation results appear highly similar, with no apparent visual differences among them; however a major difference is that the standard deviations of the offset vectors decrease by $\sim$1/5 when the absolute loss term is included in addition to the double-difference loss (Table 4), which is comparable to the ratio obtained for the California case (Table 2). 

When only using the double-difference loss, the average double-difference residuals decrease from 0.90 s to 0.68 s, while they decrease from 0.90 s to 0.71 s when using both the double-difference and absolute loss. The density factors are nearly constant for each case (Table 4), which implies that including the absolute loss (Eq. 4b) in addition to the double difference loss (Eq. 3) only slightly degrades the relative relocation accuracy. It provides the benefit, however, of removing the ambiguity in the absolute travel time reference frame that appears when only using the double-difference residuals. Reducing the non-uniqueness of the inversion and the tendency for clusters to `drift' during relocation also reduces the per-event standard deviations of the relocation updates (Table 4). The model trained using the `memory state' mechanism obtains a similar spatial pattern of relocated events as the other models (Fig. 7d), while also having loss values that converge nearly $\sim$10$\times$ faster (Fig. S8). This model also has the lowest standard deviation of relocation offset vectors despite only using the double-difference residual loss; however, there are possible signs of instability in the minimization process for the memory state model that manifest as increased long-period noise in the loss function (Fig. S8).

\subsection{Northern Chile: Application to a Large Catalog of $>$750,000 events in a Subduction Environment}

To assess the ability of our model to be applied to a very large, noisy catalog, we apply the technique to the $\sim$750,000 event catalog from 2010$-$2018 in northern Chile, derived in \citet{mcbrearty2019earthquake}. This catalog used an energy-based picker similar to STA/LTA, and a back-projection-based association scheme, so many events have poor pick times, and in some cases, poor associations as well. Furthermore, during the initial location of the events, prior information was used to encourage events to follow the subduction interface of the Crust 1.0 Slab model for deeper events \citep{laske2012crust1}, and a 3D velocity model derived from a local 2D trench perpendicular model was used for location \citep{comte1994velocity}. 

These choices differ from an established double-difference relocated catalog of northern Chile \citep{sippl2018seismicity, sippl2023northern}, which has $\sim$108,000 events for this interval, and which used a different velocity model and location scheme. As such, the initial locations of \citet{mcbrearty2019earthquake} (Figs. 8a,d) do not define the shallow faults as clearly at less than 50 km depth, and show significant bias in depth for deeper events $\sim$80$-$250 km compared to the \citet{sippl2023northern} catalog (Figs. 8c,f). To facilitate easier comparison and reduce the systematic spatial differences between the two catalogs, we train this model using the double-difference residual loss (Eq. 3) with both calibration and station correction losses (Eq. 4), weighted with a ratio of 4 : 1. This enables mapping the initial catalog closer to the absolute reference frame of \citet{sippl2023northern} while also minimizing the double-difference residuals and improving the precision of the event locations.

The relocated catalog we obtain significantly improved the resolution of shallow crustal faults (Figs. 8b,e), which is comparable with the resolution of the \citet{sippl2023northern} catalog (Figs. 8c,f). In addition, we resolve the double-seismic zone of the subducting slab (Fig. 8e), and recover a number of dense clusters related to quarry and mine activities \citep{sippl2023northern}. During training, the loss converges smoothly (Fig. 1), the density factor increases by 65$\%$, and the average double-difference residuals decrease from 0.97 s to 0.61 s (Table 2). The differences of locations between our catalog and the \citet{sippl2023northern} catalog have standard deviations of $\sim$0.06$^{\circ}$ and $\sim$0.1$^{\circ}$ for latitude and longitude, respectively, and the depth and origin time residuals reduce from 9 $\pm$ 16.4 km to -0.26 $\pm$ 12.9 km, and from 1.3 $\pm$ 1.53 s to -0.34 $\pm$ 1.08 s (Table 3). This highlights the reduction in bias and improved consistency between the two catalogs, despite the substantially different methods used to create them. 

A bias in depth is still apparent for intermediate-depth events $>$80 km; however, the slope of the down-going slab is much more consistent. The \citet{sippl2023northern} catalog is also very sparse in this region compared to the \citet{mcbrearty2019earthquake} catalog, and so only a minimal amount of `calibration' is possible for deep events. Due to this sparsity, the estimated station corrections $\Delta \vect{s}$ are likely to have been dominated by the influence of shallow events. We also locate a larger proportion of events above the subducting slab for distances $>$250 km from the trench (Fig. 8e). These likely reflect the removal of the priors influence in constraining the locations of the \citet{mcbrearty2019earthquake} catalog to the Crust 1.0 Slab model, and the small number of \citet{sippl2023northern} events that occupy this volume and are used for calibration between the two catalogs. When the model is trained using only the double-difference objective, similar resolution is maintained as when including the absolute and calibration objectives (Fig. 9). So while it is convenient to calibrate against the \citet{sippl2023northern} catalog, this is not required for obtaining high precision locations.

\subsection{Synthetic Test on an Extremely Large Catalog}

We implement a simple synthetic test to demonstrate that under idealized conditions, we can recover near perfect relocation accuracy while also applying our model to a very large catalog ($>$ 1,000,000 events) over a large geographic region. Specifically, we generate a synthetic catalog by sampling $\sim$1.25 million sources randomly from the recorded NCEDC locations throughout northern California during 2023, and adding random Gaussian noise scaled by $0.01^{\circ}$ in latitude and longitude, and 1.0 km in depth. This represents the `true' source locations for which we generate arrival times on a subset of the network with a distance threshold sampled between 30$-$300 km of the source. From this we delete up to 70\% of the arrivals across the network randomly. This creates a large, spatially extended ground truth catalog with variable monitoring conditions (Fig. 10c). For the synthetic test, we randomly record the `initial location' of each event by perturbing the true location with Laplacian error scaled by $0.025^{\circ}$ in latitude and longitude, and 2.5 km in depth. This causes the initial locations to be highly diffuse and have significant offsets compared to the ground truth (Fig. 10a). 

We build the input graphs and train the relocation model using the double-difference loss (Eq. 3), for which we see a very smooth and stable reduction of the loss function (Fig. 1). The relocated catalog has an increase in the density factor by $\sim$126$\%$ and the average double-difference residual decreases from 0.58 s to 0.04 s. Compared to ground truth locations, the epicentral residuals are nearly zero (e.g., 0.0 $\pm$ 0.007$^{\circ}$; Table 3), while the depth residuals are -0.06 $\pm$ 1.10 km, corresponding to a close match with the ground truth locations (Fig. 10b). The origin time residuals, however, are -0.50 $\pm$ 0.07 s, which indicates a drift in the absolute travel time reference frame that appears when minimizing the double-difference residuals. By re-training the model using the absolute loss in addition to the double-difference loss (scaled by 1 : 4), this origin time residual is reduced to -0.02 $\pm$ 0.07 s while preserving high spatial accuracy (Table 3). This simple test highlights the importance of including the absolute loss term to stabilize the inversion, while also indicating that under idealized conditions, near perfect relocation accuracy is possible.

\section{Discussion}

An important advantage of using a GNN for double-difference relocation is that it simplifies applying relocation to very large catalogs (e.g., Figs. 9, 10); however, there are secondary benefits and insights that we gain from implementing this approach. For instance, because automatic differentiation is used to minimize the loss, it is straightforward to include additional loss function objectives such as station corrections, absolute travel time residuals, and the `calibration objective' (Eq. 4) to shift the absolute reference frame to match a reference catalog. This allowed correcting for the offset of events from the San Andreas Fault that occur due to using a 1D velocity model that does not account for lateral heterogeneity across the fault (Figs. 2,3), the correction of depth offsets at Geysers (Fig. 5), and improved consistency between our northern Chile catalog and the high-precision \citet{sippl2023northern} catalog (Fig. 8), while simultaneously improving the relative event locations. Even when including the absolute loss in addition to the double-difference loss, the relocated T\"{u}rkiye catalog obtains a similar `density factor' and decrease in the double-difference residual as when only training with the double-difference residual loss (Table 4). The absolute loss term nonetheless played an important role in reducing the standard deviations of the per-event relocation vectors (Table 4), as well as removing the small bias in origin time estimates obtained for the synthetic catalog when only training with the double-difference residual (Table 3).

An additional benefit of our approach is that because we use a travel time neural network surrogate model, which can predict travel times for 1D or 3D velocity models for any source-station location pairs, there is no inherent linearization of the travel times as in standard HypoDD \citep{waldhauser2000double}. In other words, relocation in a single update step is possible while accounting for the full 3D heterogeneity without any modifications to the approach. This feature, combined with the iterative training of the GNN over many `trial and error' predictions of the relocation vectors of events in the catalog, allows the GNN to gradually learn to predict relocation values that approach the ground truth result in a single iteration, or location update step. That is, unlike traditional HypoDD that uses multiple (linearized) updates of the catalog locations with conjugate gradient descent to converge on the relocation values, once trained, the GNN approximates the full relocation vector for all events in a single forward call. This was demonstrated in the synthetic test, where once trained the GNN minimized the double-difference residual from 0.58 s to near zero (e.g., 0.04 s), while also obtaining very accurate locations (Table 3). This feature was also demonstrated in our test in northern California, where we re-trained the GNN in a `second iteration' on the updated locations after the first training and relocation, and the resulting relocations after the second iteration were highly consistent with the first iteration's locations (Fig. S5). The relocation vector distances in the second iteration were only $\sim$10$-$15$\%$ the relocation distances of the first iteration (Figs. S6, S7). Hence, while multiple rounds of training and application are possible, and can improve precision, a single iteration is sufficient to obtain most of the improvement in location accuracy.

An important question in using this method is how sensitive the results are to hyper-parameter choices. Traditional relocation approaches have a variety of tunable parameters with often complex trade-offs in performance and convergence, and there are a number of distinct algorithms with different behavior \citep{pavlis1980mixed, waldhauser2000double, zhang2003double, trugman2017growclust, trugman2023growclust3d, lin2018source}. Although our approach requires constructing input graphs (see section `Turning Catalogs into Input Graphs'), and there are several scale-dependent and heuristic schemes employed in this step, the obtained relocations appear largely insensitive to these choices. When constructing the station graphs for northern California, whether we use 30-nearest-neighbors or 80-km radius graphs, the resulting locations are highly similar (Fig. S4). These graphs have both a widely different number of edges and spatial distribution of connectivity (Fig. S1). The lack of sensitivity to this choice indicates the observed data characteristics control the final locations more than the graph construction schemes. This is consistent with most GNN applications, where a range of graph construction schemes lead to successful results provided they have reasonable connectivity and clustering properties \citep{zhou2020graph}. In an early test, we did find it was essential to include a dense enough sample of sources for every input graph, so that enough nearby sources can simultaneously interact to minimize their relative double-difference residuals. If the random samples of input sources are chosen completely randomly from the full catalog, in our experiments this led to too diffuse a population of sources for each input to effectively enable relocation.  That is, with sparse source sampling, the power of the double difference approach is suppressed. 

Our tests indicate that Graph Double Difference (GraphDD) is a stable model that can be customized and applied to diverse relocation problems with highly different source distributions, monitoring conditions, and data quality; however, some aspects of our method could still be refined, such as: improving the construction scheme of input catalogs, balancing the relative weights of the loss terms more carefully, systematically applying the `multi-iteration' method of relocation, and improving understanding of the `memory state' mechanism. While our results indicate a lack of sensitivity to the input graph constructions, customized graph structure might lead to improved performance \citep{hamilton2017representation, deac2022expander}. Furthermore, refined relative weightings of the loss terms may be possible; while we used the L1 norm as our training objective, experimenting with the L2 norm or mixed norms \citep{huber1992robust} may lead to improved precision. Once trained, the per-event residuals and uncertainties could be inspected to detect and suppress outliers. In some experiments, by including the calibration objective we improved relocation accuracy and reduced location bias; however, notable differences remained, such as at Geysers in California (Fig. 4) and for deep events in Chile (Fig. 8). These differences could be reduced by simply using a velocity model that is more consistent with the reference catalogs. Our framework could also be adapted to allow spatially variable station corrections \citep{richards2000earthquake}, which could reduce bias and improve absolute location accuracy \citep{lomax2022high}. This would require embedding into the station graph while only `aggregating' over a subset of the source nodes, depending on the queried source coordinates, which is a technique that has been implemented in previous models \citep{mcbrearty2023earthquake}. Finally, the `memory state' variant of the model can allow very rapid convergence (Fig. S8), while providing similar, or improved relocation results (Fig. 7d), though we note this approach has not been tested as thoroughly as the default implementation. Future effort could more systematically understand the trade-off of these different training options and optimize performance on specific datasets.

\section{Conclusions}

We implement a deep learning approach for the joint relocation of an earthquake catalog, which allows joint relocation over very large catalogs ($>$1 million events), and is easily adapted to handle 1D or 3D velocity models with variable loss functions. By using the Cartesian product graph between sources and stations, and including pick residuals and travel time partial derivatives in the input, the proposed architecture is well-suited for minimizing the double-difference residual. The technique also allows merging absolute and relative arrival time residual loss terms and incorporating station corrections. This enables customized applications by specifying the importance of different location objectives, such as mapping the locations into the reference frame of a different catalog. We believe this a promising approach for hypocenter relocation, and is particularly well-suited for processing the latest generation of machine-learning-based catalogs with very large numbers of earthquakes, including many that are near the detection threshold.

\section{Data and Resources}

The catalogs used in this study include the northern Chile catalogs \citep{mcbrearty2019earthquake, sippl2023northern}, the T\"{u}rkiye catalog \citep{becker2024performance}, and an extension of the northern California catalog derived in \citep{mcbrearty2023earthquake}. We also make use of the NCEDC \citep{NCEDC_2014_7932} and AFAD catalogs \citep{AFAD2023catalog}. The locations for the initial and relocated events are provided in the supplemental materials.

\section{Declaration of Competing Interests}

The authors acknowledge there are no conflicts of interest recorded.

\section{Acknowledgments}

We thank Weiqiang Zhu for providing the PhaseNet picks used in the northern California catalog, and Albert Aguilar for help in using GMT. We also thank William Ellsworth, Xing Tan, and Yifan Yu for helpful discussions. This work was supported by AFRL under contract PA04-S-D00243-12-TO-03-Stanford.

\newpage

\bibliography{bib}
\bibliographystyle{apa}

\newpage

\begin{table} \centering

\caption{Summary of Input Graph Statistics. For each of the catalogs used in the study, we list the average statistics of the dataset and input graphs, including the total number of events in the catalog, the average number of events per input graph sample, the total number of stations, the average number of picks per input graph sample, the average number of source-source and station-station edges in the input Cartesian product subgraphs, and the latitude and longitude range considered. Results are measured by averaging outputs over 100 random input graphs.}

\begin{minipage}{1.075\linewidth}

\vspace{0.5cm}

\begin{tabularx}{\textwidth}{c c c c c c c c}
\toprule

Catalog & Number & Num. Events & Number & Num. Picks & Src-Src & Sta-Sta & Spatial \\
Name & Events & per Sample & Stations & per Sample & Edges & Edges & Region \\

\midrule
California & 71k & 8k & 835 & 157k & 947k & 2.4M & 3.8$^{\circ}\times$3.6$^{\circ}$ \\
T\"{u}rkiye & 13k & 6k & 40 & 75k & 563k & 312k & 2.3$^{\circ}\times$3.6$^{\circ}$ \\
Chile & 776k & 15k & 17 & 168k & 1.5M & 1.3M & 8.0$^{\circ}\times$5.2$^{\circ}$ \\
Synthetic & 1.2M & 3k & 914 & 174k & 705k & 2.8M & 8.2$^{\circ}\times$8.0$^{\circ}$ \\

\bottomrule
\end{tabularx}

\end{minipage}
\end{table}

\newpage

\begin{table} \centering

\caption{Summary of Relocation Results. For each of the catalogs used in the study, we list the average offset vector length (km), the standard deviation of the offset vectors (km; for a fixed source, averaged over many input graphs), the `density factor', the average double difference residual, and the average double difference residual excluding the upper $5\%$ of residual values. The density factor is defined as the mean value of the inverse of the mean distance of each event to its 50 nearest neighbors. Each row for the same catalog represents the values before (upper) and after (lower) relocation. Results are measured by averaging outputs over 1000 random input graphs.}

\begin{minipage}{0.82\linewidth}

\vspace{0.5cm}

\begin{tabularx}{\textwidth}{c c c c c c}
\toprule

  Catalog & Mean & Std. & Density & Average & Avg. Residual \\
  Name &  Offset & Offset & Factor & Residual & ($<$95$\%$ Quant.) \\
\midrule
California & - & - & 0.93 & 0.18 & 0.15 \\ 
& 5.2 & 0.5 & \textbf{3.72} & \textbf{0.13} & \textbf{0.10} \\ 
\cmidrule(lr){1-6}
 T\"{u}rkiye & - & - & 0.22 & 0.90 & 0.76 \\
& 6.2 & 5.7 & \textbf{0.32} & \textbf{0.68} & \textbf{0.52} \\
\cmidrule(lr){1-6}
Chile & - & - & 0.32 & 0.97 & 0.82 \\
& 27.4 & 3.2 & \textbf{0.53} & \textbf{0.61} & \textbf{0.44} \\
\cmidrule(lr){1-6}
Synthetic & - & - & 2.25 & 0.58 & 0.50 \\
& 4.7 & 0.3 & \textbf{5.09} & \textbf{0.04} & \textbf{0.03} \\

\bottomrule
\end{tabularx}

\end{minipage}
\end{table}

\newpage

\begin{table} \centering

\caption{Summary of Relocation Comparisons to Reference Catalogs. For each of the catalogs used in the study, we list the average latitude, longitude, depth, and origin time offsets to the reference catalogs both before (upper) and after (lower) relocation. For the synthetic catalog we list the result using only the double difference residual loss, and using the double difference and absolute travel time residual loss.}

\begin{minipage}{1.05\linewidth}

\vspace{0.5cm}

\begin{tabularx}{\textwidth}{c c c c c}
\toprule

  Catalog & Lat. & Lon. & Depth & Origin \\
  & Offset & Offset & Offset (km) & Offset (s) \\
\midrule
California & -0.004 ($\pm$ 0.018) & -0.003 ($\pm$ 0.020) & -4.30 ($\pm$ 3.90) & -0.53 ($\pm$ 0.58) \\ 
& \textbf{0.00 ($\pm$ 0.010)} & \textbf{-0.001 ($\pm$ 0.016)} & \textbf{-0.52 ($\pm$ 2.30)} & \textbf{0.04 ($\pm$ 0.40)} \\ 
\cmidrule(lr){1-5}
 T\"{u}rkiye & 0.017 ($\pm$ 0.104) & \textbf{0.008 ($\pm$ 0.144)} & 3.60 ($\pm$ 5.40) & \textbf{-0.29 ($\pm$ 1.50)} \\
& \textbf{-0.001 ($\pm$ 0.100)} & 0.040 ($\pm$ 0.141) & \textbf{-0.58 ($\pm$ 4.80)} & -0.38 ($\pm$ 1.49) \\
\cmidrule(lr){1-5}
Chile & -0.002 ($\pm$ 0.073) & -0.061 ($\pm$ 0.115) & 9.00 ($\pm$ 16.40) & 1.30 ($\pm$ 1.53) \\
& \textbf{-0.002 ($\pm$ 0.061)} & \textbf{0.049 ($\pm$ 0.105)} & \textbf{-0.26 ($\pm$ 12.90)} & \textbf{-0.34 ($\pm$ 1.08)} \\
\cmidrule(lr){1-5}
Synthetic & 0.00 ($\pm$ 0.035) & 0.00 ($\pm$ 0.034) & -0.07 ($\pm$ 3.30) & \textbf{0.00 ($\pm$ 0.18)} \\
& \textbf{0.00 ($\pm$ 0.006)} & \textbf{0.00 ($\pm$ 0.007)} & \textbf{-0.06 ($\pm$ 1.10)} & -0.50 ($\pm$ 0.07) \\
\cmidrule(lr){1-5}
Synthetic & 0.00 ($\pm$ 0.035) & 0.00 ($\pm$ 0.034) & -0.07 ($\pm$ 3.30) & 0.00 ($\pm$ 0.18) \\
(w/ abs) & \textbf{0.00 ($\pm$ 0.007)} & \textbf{0.00 ($\pm$ 0.008)} & \textbf{-0.10 ($\pm$ 1.20)} & \textbf{-0.02 ($\pm$ 0.07)} \\

\bottomrule
\end{tabularx}

\end{minipage}
\end{table}

\newpage

\begin{table} \centering

\caption{Summary of Relocation Results for T\"{u}rkiye, using different variants of the loss function. We consider the double difference residual loss, both the double difference residual and absolute travel time residuals, the double difference and absolute travel time residuals with station corrections, and the model trained with the double difference residual loss using the `memory state' mechanism. For each variant, we list the average offset vector length (km), the standard deviation of the offset vectors (km; for a fixed source, averaged over many input graphs), the `density factor', the average double difference residual, and the average double difference residual excluding the upper $5\%$ of residual values. Each row for the same catalog represents the values before (upper) and after (lower) relocation. Results are measured by averaging outputs over 1000 random input graphs. For the models trained with multiple loss values, we weight the double difference loss value with 4/5 and the additional loss values with 1/5 during training.}

\begin{minipage}{0.82\linewidth}

\vspace{0.5cm}

\begin{tabularx}{\textwidth}{c c c c c c}
\toprule

  Catalog & Mean & Std. & Density & Average & Avg. Residual \\
  Name &  Offset & Offset & Factor & Residual & ($<$95$\%$ Quant.) \\
\midrule
T\"{u}rkiye & - & - & 0.22 & 0.90 & 0.76 \\
(regular) & 6.2 & 5.7 & \textbf{0.32} & \textbf{0.68} & \textbf{0.52} \\
\cmidrule(lr){1-6}
 T\"{u}rkiye & - & - & 0.22 & 0.90 & 0.76 \\
(memory) & 7.9 & 0.42 & \textbf{0.32} & \textbf{0.67} & \textbf{0.51} \\
\cmidrule(lr){1-6}
T\"{u}rkiye & - & - & 0.22 & 0.90 & 0.76 \\
(w/ absolute) & 5.7 & 1.0 & \textbf{0.31} & \textbf{0.70} & \textbf{0.54} \\
\cmidrule(lr){1-6}
T\"{u}rkiye & - & - & 0.22 & 0.90 & 0.76 \\
(w/ abs and $\Delta$s) & 4.8 & 1.1 & \textbf{0.31} & \textbf{0.71} & \textbf{0.55} \\

\bottomrule
\end{tabularx}

\end{minipage}
\end{table}

\begin{figure}[ht!] 
    \centering
    \includegraphics[width=1.0\textwidth]{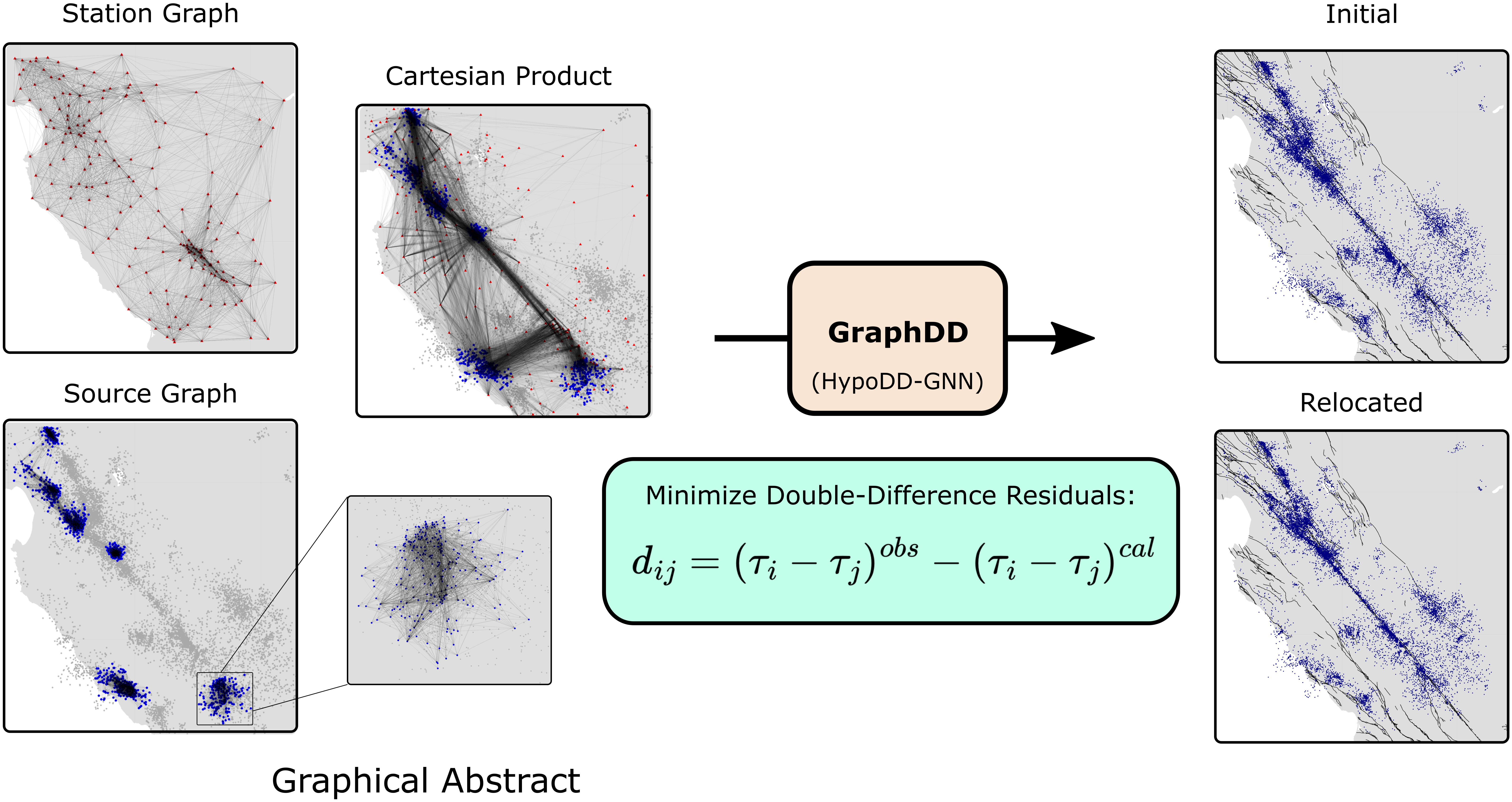}
\end{figure}

\setcounter{figure}{0}

\begin{figure}[ht!] 
    \centering
    \includegraphics[width=0.75\textwidth]{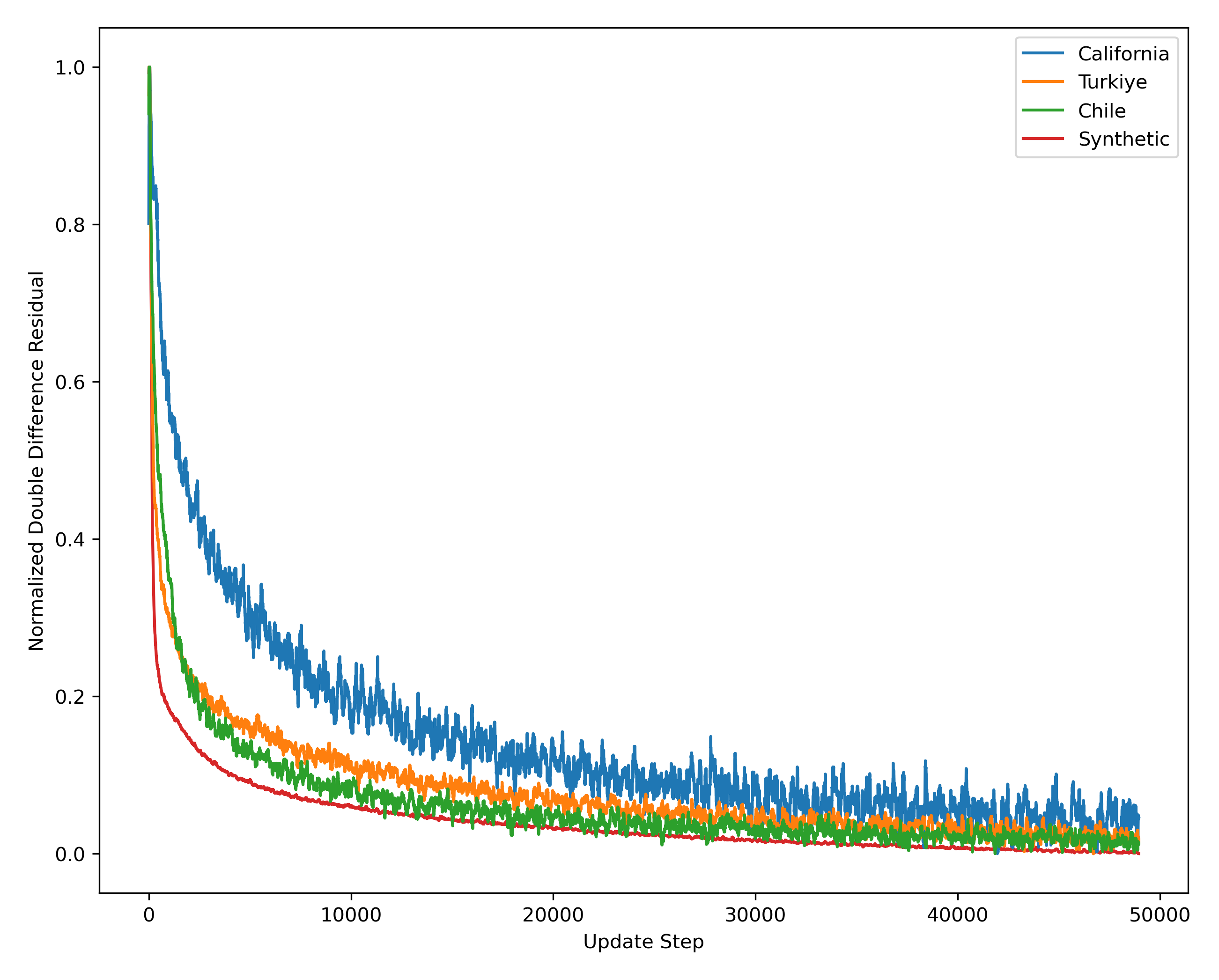}
    \caption{The double difference residual loss function of the trained model for the California, T\"{u}rkiye, Chile catalogs and the synthetic test. The loss curves are normalized to the same [0,1] range for all models for easier visual comparisons and are plotted as the moving median over 100 sequential update steps.}
\end{figure}

\begin{figure}[ht!] 
    \centering
    \includegraphics[width=0.9\textwidth]{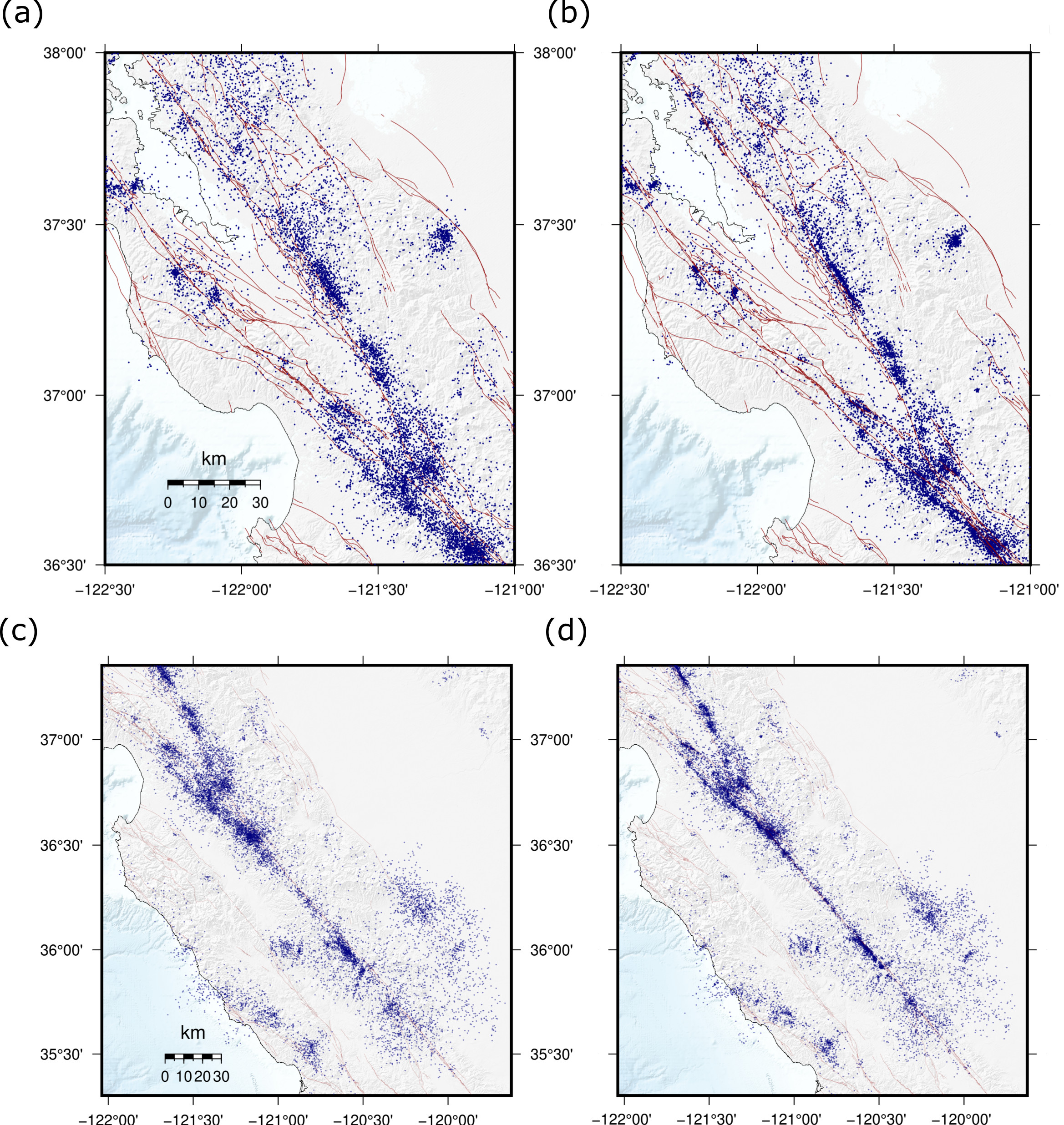}
    \caption{The initial (a,c) and relocated (b,d) northern California catalog, for the north region (a,b) and the southern region (c,d).}
\end{figure}

\begin{figure}[ht!] 
    \centering
    \includegraphics[width=0.9\textwidth]{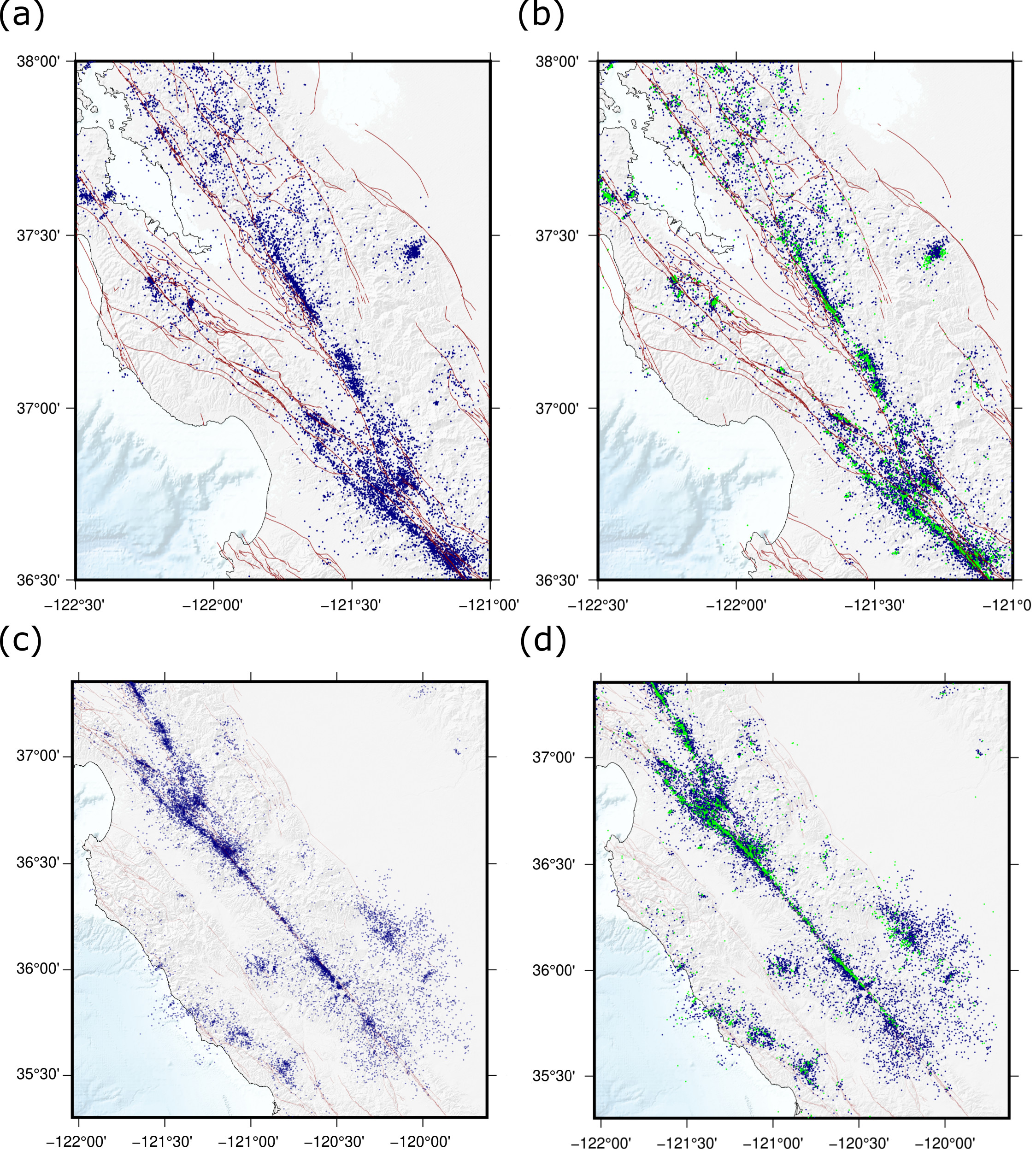}
    \caption{The relocated northern California catalog (a,c) and the relocated catalog with the NCEDC catalog overlain in green (b,d).}
\end{figure}

\begin{figure}[ht!] 
    \centering
    \includegraphics[width=1.0\textwidth]{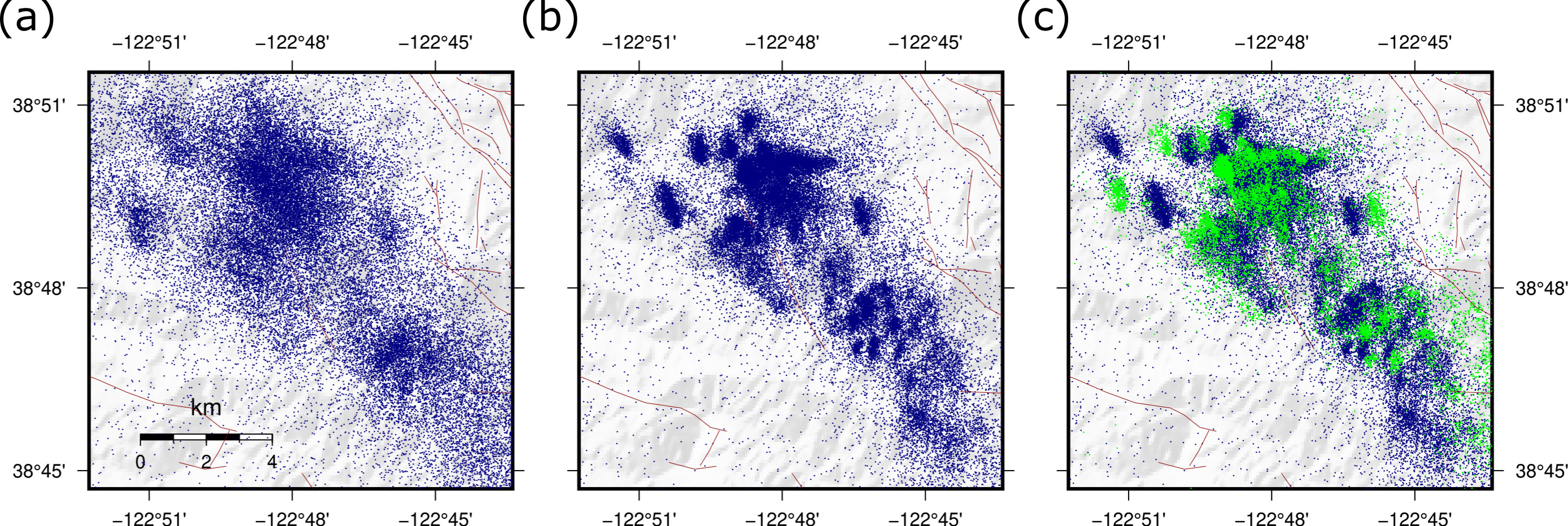}
    \caption{The initial (a), relocated (b), and relocated catalog with the NCEDC catalog overlain in green (c) for the Geysers geothermal field.}
\end{figure}

\begin{figure}[ht!] 
    \centering
    \includegraphics[width=0.8\textwidth]{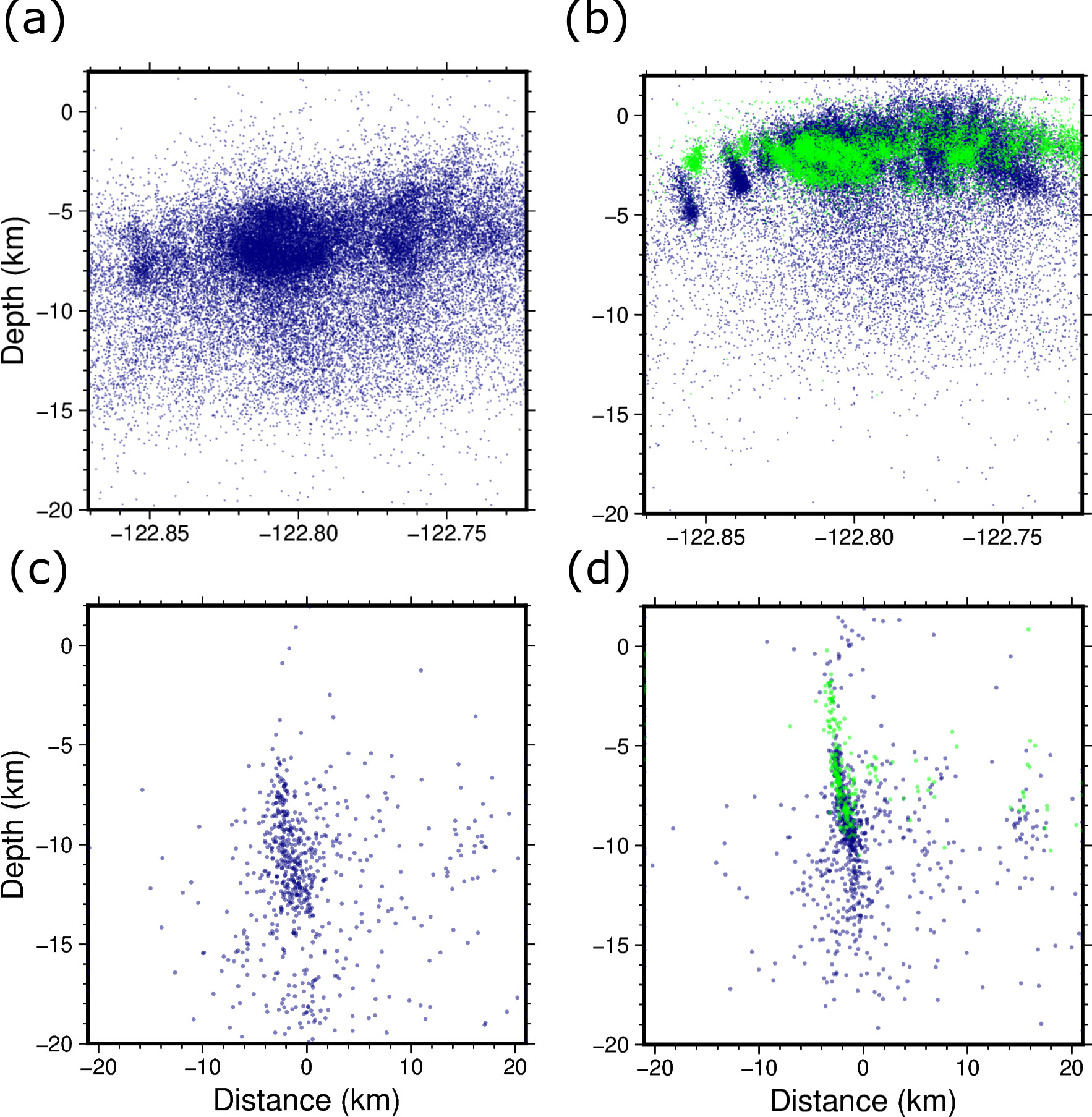}
    \caption{Cross-sections of the initial (a,c) and relocated catalog with the NCEDC overlain in green (b,d). The transects include the Geysers geothermal field (a,b) and the Calaveras fault (c,d), and all events within $\pm$0.1$^{\circ}$ of the transect.}
\end{figure}

\begin{figure}[ht!] 
    \centering
    \includegraphics[width=0.9\textwidth]{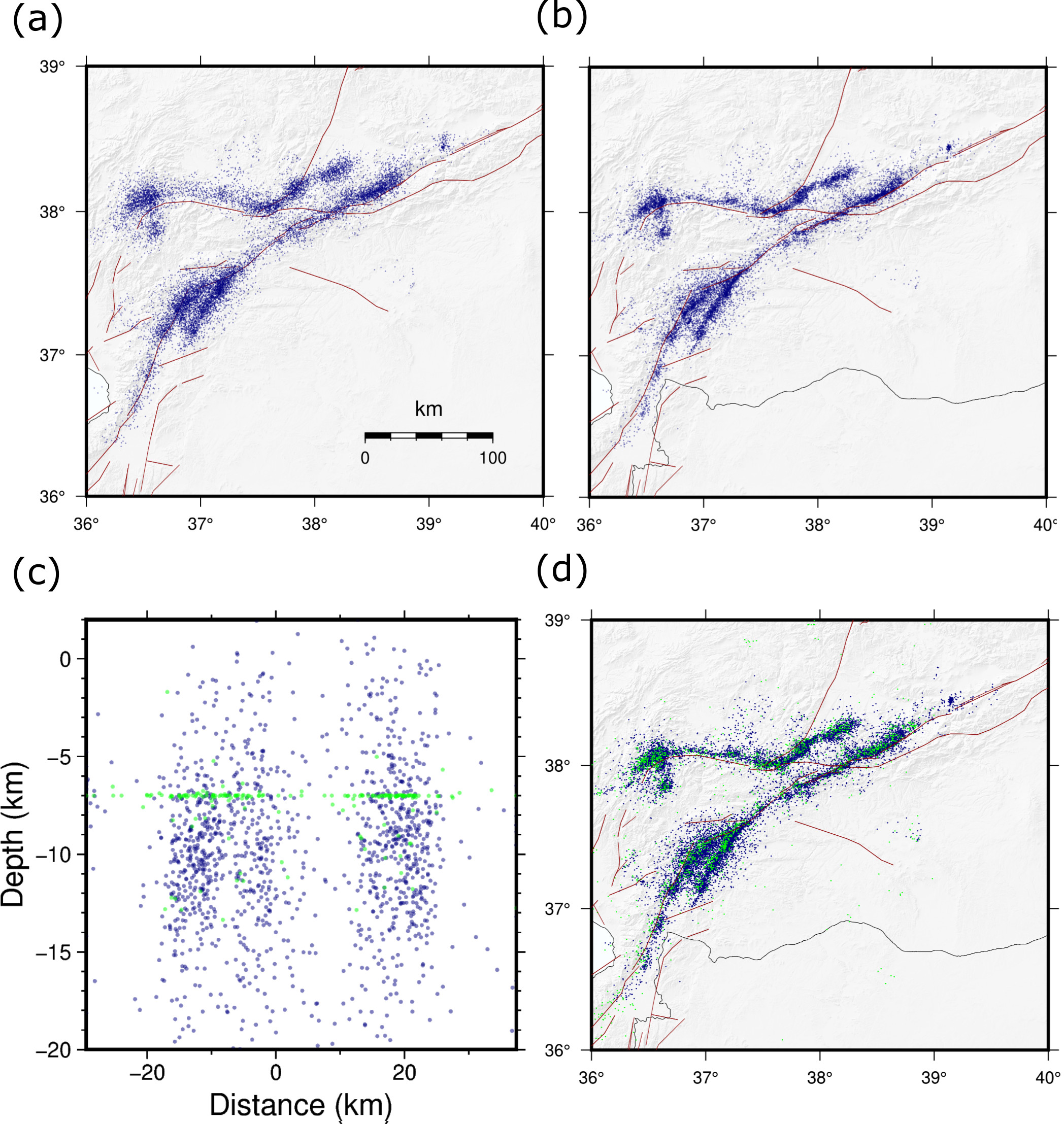}
    \caption{The initial (a) and relocated (b) T\"{u}rkiye catalog. Depth cross-sections across the \c{C}ardak fault and EAFZ are shown in (c), and the relocated catalog along with the AFAD catalog overlain in green are shown in (d).}
\end{figure}

\begin{figure}[ht!] 
    \centering
    \includegraphics[width=1.0\textwidth]{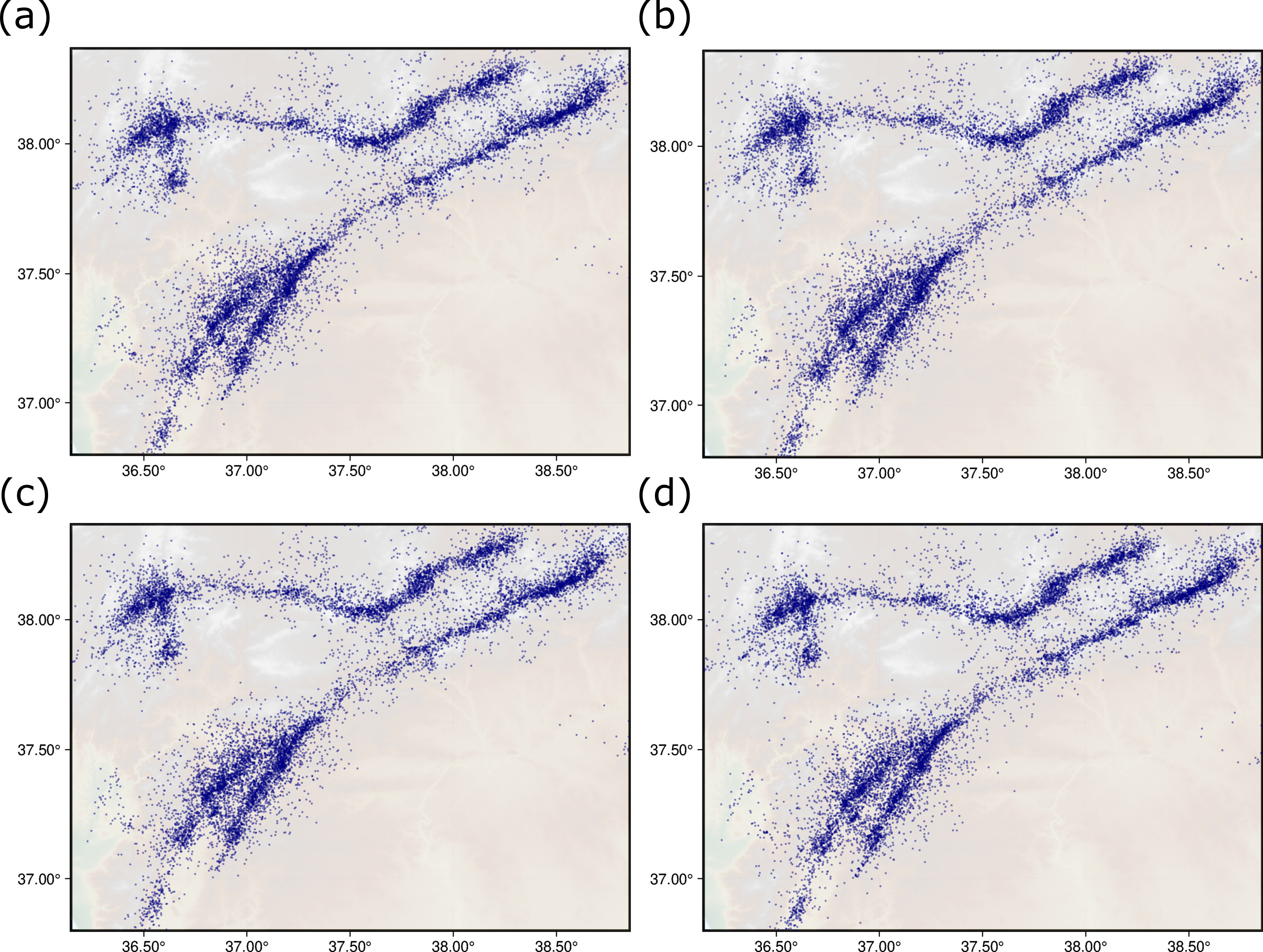}
    \caption{A comparison of the relocated T\"{u}rkiye catalog obtained using (a) the double-difference residual loss, (b) the double-difference with absolute loss, (c) the double-difference and absolute losses with station corrections, (d) and the double-difference loss trained with the `memory state' variant of the model.}
\end{figure}

\begin{figure}[ht!] 
    \centering
    \includegraphics[width=1.1\textwidth]{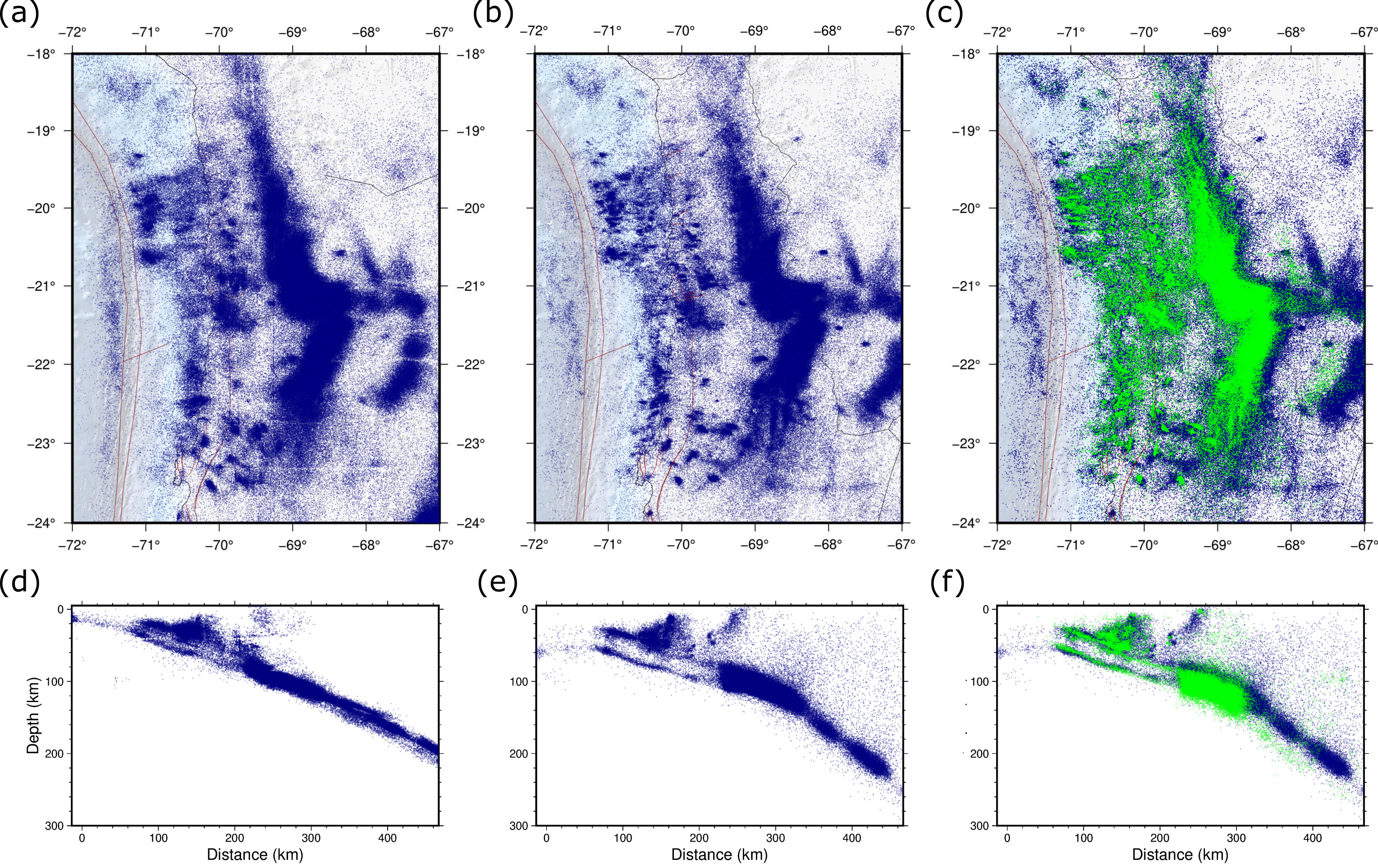}
    \caption{The initial (a) and relocated (b) northern Chile catalog, and the relocated catalog with the \citet{sippl2023northern} catalog overlain in green (c). The cross-sections (d-f) are plotted at 21.25$^{\circ}$S, with distance measured from the trench, and include all events within $\pm 0.25$$^{\circ}$S of the corresponding catalogs shown in (a-c).}
\end{figure}

\begin{figure}[ht!] 
    \centering
    \includegraphics[width=1.0\textwidth]{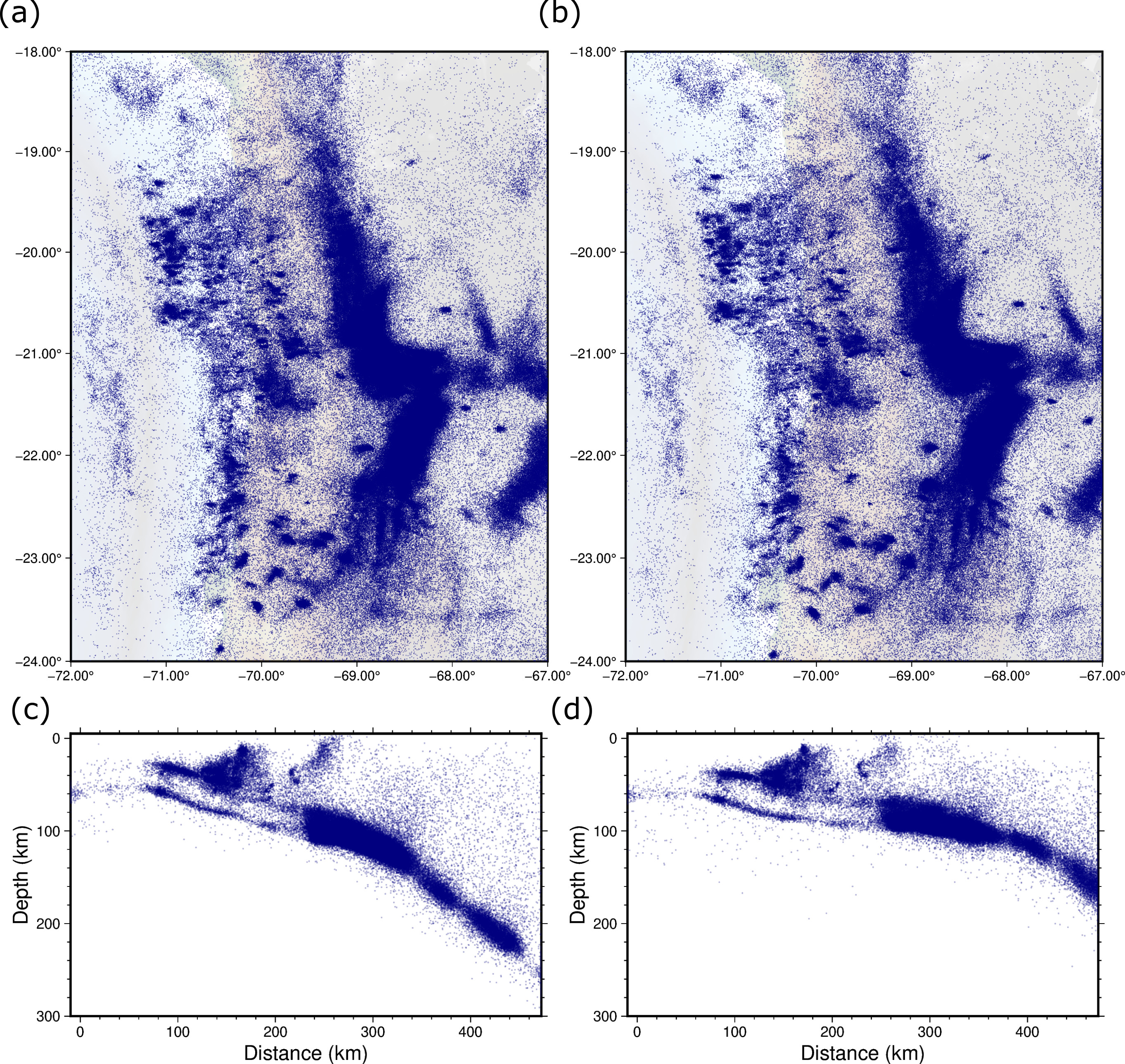}
    \caption{A comparison of the relocated northern Chile catalog obtained using (a) the double-difference, absolute, and calibration objectives, and (b) only the double-difference objective. The cross-sections (c,d) are plotted at 21.25$^{\circ}$S, with distance measured from the trench, and include all events within $\pm 0.25$$^{\circ}$S of the corresponding catalogs shown in (a,b).}
\end{figure}

\begin{figure}[ht!] 
    \centering
    \includegraphics[width=1.0\textwidth]{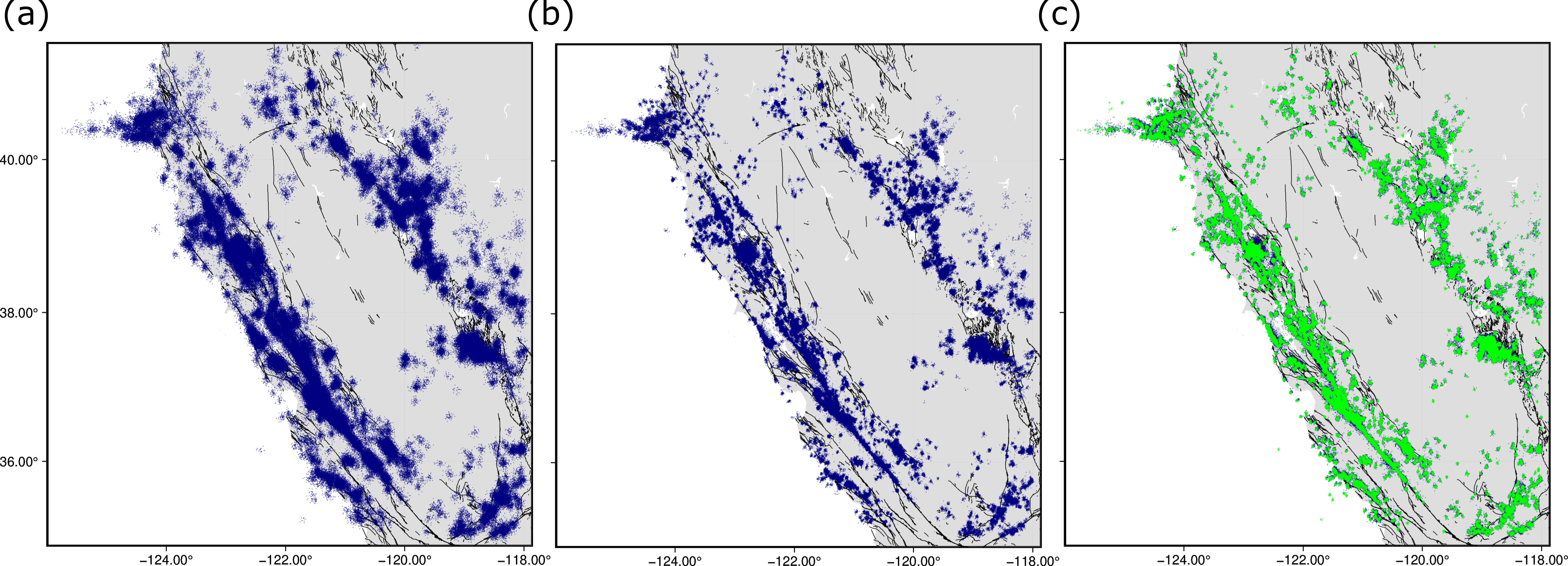}
    \caption{The initial (a) and relocated (b) synthetic northern California catalog, and the relocated catalog with the ground truth locations overlain in green (c).}
\end{figure}

\end{document}